\documentclass[prl,aps,showpacs,twocolumn,unsortedaddress,superscriptaddress]{revtex4}
\usepackage{graphics,bm}
\usepackage{amssymb,ulem}
\usepackage{epsfig}
\usepackage{epsf}
\usepackage[usenames]{color}

\begin{document}

\title{Spin drag in an ultracold Fermi gas on the verge of a ferromagnetic instability}

\author{R.A. Duine}
\affiliation{Institute for Theoretical Physics, Utrecht
University, Leuvenlaan 4, 3584 CE Utrecht, The Netherlands}

\author{Marco Polini}
\affiliation{NEST Istituto di Nanoscienze-CNR and Scuola Normale Superiore,
I-56126 Pisa, Italy}

\author{H.T.C. Stoof}

\affiliation{Institute for Theoretical Physics, Utrecht
University, Leuvenlaan 4, 3584 CE Utrecht, The Netherlands}

\author{G. Vignale}
\affiliation{Department of Physics and Astronomy, University of
Missouri, Columbia, Missouri 65211, USA}

\date{\today}

\begin{abstract}
Recent experiments [Jo {\it et al.}, Science {\bf 325}, 1521
(2009)] have presented evidence of ferromagnetic correlations in a
two-component ultracold Fermi gas with strong repulsive
interactions. Motivated by these experiments we consider spin
drag, {\it i.e.}, frictional drag due to scattering of particles
with opposite spin, in such systems. We show that when the
ferromagnetic state is approached from the normal side, the spin
drag relaxation rate is strongly enhanced near the critical point.
We also determine the temperature dependence of the spin diffusion
constant. In a trapped gas the spin drag relaxation rate
determines the damping of the spin dipole mode, which therefore
provides a precursor signal of the ferromagnetic phase transition
that may be used to experimentally determine the proximity to the
ferromagnetic phase.
\end{abstract}


\maketitle

\def\bx{{\bm x}}
\def\bk{{\bm k}}
\def\bK{{\bm K}}
\def\bq{{\bm q}}
\def\br{{\bm r}}
\def\bp{{\bm p}}
\def\half{\frac{1}{2}}
\def\args{(\bm, t)}

{\it Introduction.} --- Interest in electronic transport ranges
from everyday applications to fundamental physics. One of the most
interesting phenomena that spans this entire range, is the
influence of a thermodynamic phase transition on the electrical
conductivity. The most direct example is the phase transition from
a normal conductor to a superconductor characterized by a
vanishing resistance. The applications of this phenomenon are
ubiquitous and the basic physics that underlies the transition in
superconductors, the Bose-Einstein condensation of fermionic
pairs, has emerged in research fields from astroparticle
physics~\cite{son1999} to cold-atom
systems~\cite{stoof1996,regal2004}.

\begin{figure}[t]
\begin{center}
\includegraphics[width=1.00\linewidth]{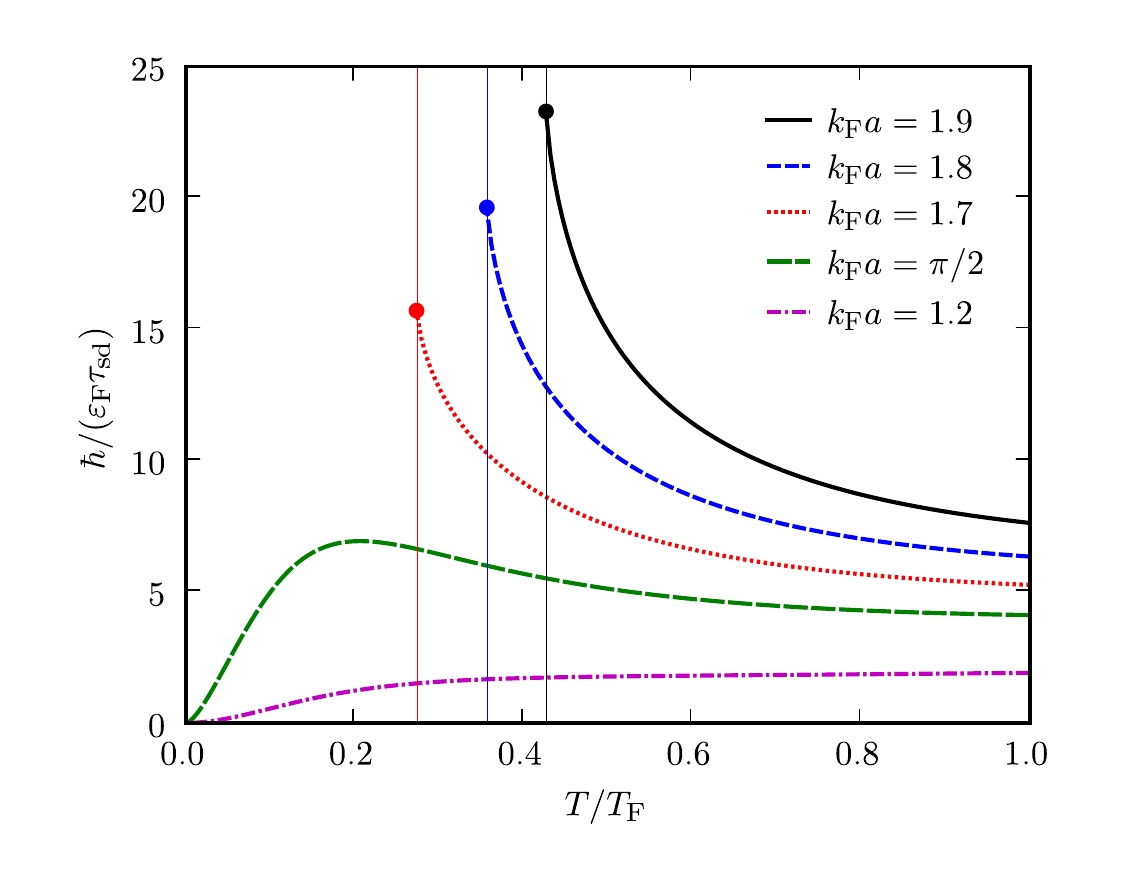}
\caption{(Color online) Spin drag relaxation rate $1/\tau_{\rm
sd}(T)$ as a function of temperature $T$, for various values of
the interaction parameter $k_{\rm F} a$. The Fermi energy is
denoted by $\varepsilon_{\rm F} = k_{\rm B}T_{\rm F} = \hbar^2
k_{\rm F}^2/2m$. Note that for $k_{\rm F}a > \pi/2$, the spin drag
relaxation rate shows a distinctive upturn when the critical
temperatures, indicated by thin vertical lines, are approached
from above. In particular, we have $T_{\rm c} \simeq 0.43~T_{\rm
F}$ for $k_{\rm F}a = 1.9$, $T_{\rm c} \simeq 0.36~T_{\rm F}$ for
$k_{\rm F}a = 1.8$, and $T_{\rm c} \simeq 0.27~T_{\rm F}$ for
$k_{\rm F}a = 1.7$. For $k_{\rm F} a < \pi/2$ no ferromagnetism
occurs within mean-field theory. In that case $1/\tau_{\rm sd}(T)$
is smooth throughout the temperature range and exhibits the
standard Fermi-liquid behavior $1/\tau_{\rm sd}(T) \propto T^2$
for $T \to 0$.\label{fig:tausdvsT}}
\end{center}
\end{figure}

A system in between these two temperature extremes, in which
analogies of superconductivity have been predicted, is that of a
two-dimensional (2D) electron-hole
bilayer~\cite{shevchenko1976,lozovik1976}. In this case the pairs
that condense are excitons formed by electrons from one layer with
holes in the other. The relevant transport probe is in this case
the Coulomb drag measurement~\cite{coulombdrag}: a current $I$ is
driven through one layer, known as the ``active" layer, causing a
voltage drop $V_{\rm D}$ in the other. As the layers are separated
by an essentially impenetrable tunnel barrier, the voltage drop is
predominantly caused by Coulomb scattering, and the drag
resistivity $\rho_{\rm D} = V_{\rm D} / I$ has the characteristic
quadratic Fermi-liquid-like low-temperature dependence $\rho_{\rm
D} \propto T^2$. When the excitons undergo Bose-Einstein
condensation, however, the drag resistivity is predicted to jump
from the relatively small value proportional to $T^2$ to a value
equal to the ordinary resistivity of the active
layer~\cite{vignale1996}. Although conclusive evidence of exciton
condensation is still lacking, two experimental
groups~\cite{croxall2008,seamons2009} have recently reported the
observation of an upturn in the drag resistivity as the
temperature is lowered. This upturn is interpreted as being due to
strong pairing fluctuations that precede exciton
condensation~\cite{hu2000} and thus serves as a precursor signal
for the transition, similar to the enhancement of the conductivity
in superconductors due to superconducting fluctuations above but
close to the critical temperature~\cite{larkin2002}.

A closely related situation arises when the two layers of a 2D electron-electron bilayer are close enough to
allow the establishment of interlayer coherence~\cite{eisenstein_nature_2004}. In this case, the
two layers in the system can be labelled ``up" and ``down" along a
``z"-axis, so that the which-layer degree of freedom becomes a
spin one-half pseudospin. Interlayer coherence in this language
corresponds to pseudospin ferromagnetism with an easy x-y plane,
since this orientation of the pseudospin describes a particle that
is neither in the left nor in the right layer, but in a coherent
superposition of the two. Furthermore, Coulomb drag becomes
pseudospin drag, the mutual friction between two pseudospin states
due to Coulomb scattering. This analogy prompted studies of spin
drag, the frictional drag between electrons with opposite spin
projection, in a single semiconductor~\cite{scd_giovanni}. While
the realization of separate electric contacts to the two spin
states remains an experimentally challenging problem, the spin
drag is observed indirectly, by measuring different diffusion
constants for charge and spin~\cite{weber_nature_2005}.

Because of the presence of other relaxation mechanisms, spin drag
effects are usually not very large in semiconductors, and are even
smaller in metals. This is completely different in cold atomic
gases where scattering between different hyperfine spins is the
only mechanism to relax spin currents, and was considered both for
fermionic atoms~\cite{polini_prl_2007,bruun_prl_2008}, and for
bosonic ones~\cite{duine_prl_2009}. It is the purpose of this
Letter to point out that a particularly interesting situation
occurs when spin drag is considered in a two-component Fermi gas
that is close to a ferromagnetic
instability~\cite{houbiers1997,salasnich2000,amoruso2000,sogo2002,duine2005}.
Based on the analogy between electron-hole bilayers and pseudospin
ferromagnets we expect that the spin drag will be strongly
enhanced as the ferromagnetic state is approached from the normal
side. Because atoms are neutral, the relevant experimental
quantity is the spin drag relaxation rate, which for instance
determines the damping rate of the spin dipole mode in trapped
cold-atom systems~\cite{bruun_prl_2008} and is thus accessible experimentally. Interestingly, an electronic analog of
the spin dipole mode also exists~\cite{damico_prb_2006}.

Our main findings are illustrated in Fig.~\ref{fig:tausdvsT}. This
plot shows the spin drag relaxation rate $1/\tau_{\rm sd} (T)$ as
a function of temperature, for various interaction strengths
determined by the product of the Fermi wave vector $k_{\rm F}$ and
the scattering length $a$. The dramatic enhancement of the
relaxation rate upon approaching the critical temperature for the
ferromagnetic transition is clearly visible. One of our
motivations for considering this effect is the recent observation
of ferromagnetic correlations in a two-component Fermi gas with
strong repulsive interactions~\cite{jo2009}. The fact that
spin-polarized domains were not directly observed adds to the
theoretical interest~\cite{conduit2009} in this experiment. The
enhancement of the spin drag relaxation rate as the ferromagnetic
phase is approached serves as a precursor probe for ferromagnetism
that is distinct from, and adds to, the experimental methods of Jo
{\it et al.}~\cite{jo2009}, and is also interesting in its own
right. In the following we present our calculations in detail, and
present additional results and discussion.

{\it Spin drag relaxation rate.} --- We consider a 3D homogeneous gas of fermionic atoms of mass $m$,
with two hyperfine states denoted by $|\uparrow\rangle$ and
$|\downarrow\rangle$. The grand-canonical Hamiltonian with
chemical potential $\mu$ is given by
\begin{eqnarray}
{\hat H} &=& \int\!d^3\bx \sum_{\alpha\in\{\uparrow,\downarrow\}}
{\hat \psi}^\dagger_\alpha (\bx)\left(- \frac{\hbar^2 {\nabla^2_\bx}}{2m} -\mu \right) {\hat \psi}_\alpha (\bx) \nonumber \\
&+& U \int\!d^3\bx~{\hat \psi}^\dagger_\uparrow (\bx) {\hat
\psi}^\dagger_\downarrow (\bx) {\hat \psi}_\downarrow (\bx) {\hat
\psi}_\uparrow(\bx)~,
\end{eqnarray}
in terms of fermionic creation and annihilation operators ${\hat
\psi}_\alpha^\dagger (\bx)$ and ${\hat \psi}_\alpha (\bx)$,
respectively. At low temperatures $s$-wave scattering, described
by $U = 4 \pi a \hbar^2/m$, dominates, and we have therefore
omitted other interaction terms from this Hamiltonian.

We first determine a frequency and momentum dependent scattering
amplitude $A_{\uparrow\downarrow} (q,\omega)$ that takes into
account many-body effects on the scattering of atoms with opposite
spin. We use the common random-phase approximation that consists
of summing all ``bubble" diagram contribution to this effective
interaction. This takes into account modifications of the
interaction due to density and spin fluctuations in an approximate
way. The latter are essential as the spin fluctuations are
strongly enhanced close to the ferromagnetic phase transition. In
terms of the noninteracting (Lindhard) response
function at nonzero temperature
\begin{equation}
\label{eq:lindhard} \chi_0(q,\omega) =2 \int \frac{d^3\bk}{(2\pi)^3}
\frac{N_{\bq+\bk}-N_\bk} {\varepsilon_{\bq+\bk}- \varepsilon_{\bk}
-\hbar\omega -i0}~,
\end{equation}
with $\varepsilon_\bk=\hbar^2{\bk}^2/2m$ and
$N_\bk=[e^{(\varepsilon_\bk-\mu)/k_{\rm B} T}+1]^{-1}$ the
Fermi-Dirac distribution function, the scattering amplitude reads
\begin{eqnarray}\label{eq:scattering_amplitude_transparent}
A_{\uparrow\downarrow}(q,\omega) = U +
\frac{U^2}{4}~\chi_{nn}(q,\omega) - \frac{3
U^2}{4}~\chi_{S_zS_z}(q,\omega)~,
\end{eqnarray}
where $\chi_{S_zS_z(nn)} (q,\omega) =  \chi_0(q,\omega)/[1\pm
U\chi_0(q,\omega)/2]$. In this notation $\chi_{nn} (q,\omega)$ is
the density-density response function while $\chi_{S_zS_z}
(q,\omega)$ describes the spin-spin response. The factor of three
in the last term in the right-hand side of
Eq.~(\ref{eq:scattering_amplitude_transparent}) comes about
because longitudinal and transverse spin fluctuations are both
taken into account.

Within Stoner mean-field theory (MFT), ferromagnetism occurs when
$\chi_{S_zS_z} (0,0)$ diverges so that $1+U \chi_0(0,0)/2=0$. This
equation gives, together with the equation $n = 2 \int
d^3\bq~N_\bq/(2\pi)^3$ for the total density determining the
chemical potential, the critical temperature $T_{\rm c}$ as a
function of $k_{\rm F} a$. For $k_{\rm B} T_{\rm c}$ much smaller
than the Fermi energy $\varepsilon_{\rm F} \equiv \hbar^2k^2_{\rm
F}/2m=\hbar^2 (3\pi^2n)^{2/3}/2m$, this gives a critical
temperature
\begin{equation}\label{eq:Tc-smallT}
 \frac{k_{\rm B} T_{\rm c}}{\varepsilon_{\rm F}} \simeq
 \frac{2\sqrt{3}}{\pi} \sqrt{\frac{U\nu (\varepsilon_{\rm F})}{2}-1} = \frac{2\sqrt{3}}{\pi} \sqrt{\frac{2k_{\rm F} a}{\pi}-1}~,
\end{equation}
where $\nu (\varepsilon_{\rm F}) = mk_{\rm F}/\pi^2\hbar^2$ is the density of
states at the Fermi level. Note that one needs $k_{\rm F} a
> \pi/2$ for the critical temperature to be nonzero, and that
there is a quantum critical point when $k_{\rm F} a =
\pi/2$~\cite{houbiers1997}.

Our next step is to use the scattering amplitude
$A_{\uparrow\downarrow} (q,\omega)$ in
Eq.~(\ref{eq:scattering_amplitude_transparent}) in the well-known
expression for the spin drag relaxation rate $1/\tau_{\rm sd}$,
following from Boltzmann
theory~\cite{scd_giovanni,bruun_prl_2008}. This yields the result
\begin{eqnarray}\label{eq:resulttausd}
\frac{1}{\tau_{\rm sd}(T)} &=& \frac{\hbar^2}{4 m n k_{\rm B} T}
\int \frac{d^3\bq}{(2\pi)^3} \frac{q^2}{3} \nonumber \\ &\times&
\int_{-\infty}^{+\infty} \frac{d\omega}{\pi}
|A_{\uparrow\downarrow}(q,\omega)|^2\frac{[\Im
m~\chi_0(q,\omega)]^2} {\sinh^2[\hbar \omega/(2 k_{\rm B} T)]},~~
\end{eqnarray}
which, together with
Eq.~(\ref{eq:scattering_amplitude_transparent}), is the central
result of this Letter and will be evaluated next. Before
proceeding we note that our result is similar to the theory of
spin diffusion in liquid He$^3$~\cite{rice1967}, although the
ferromagnetic phase transition was not considered in this context.

\begin{figure}
\begin{center}
\includegraphics[width=1.00\linewidth]{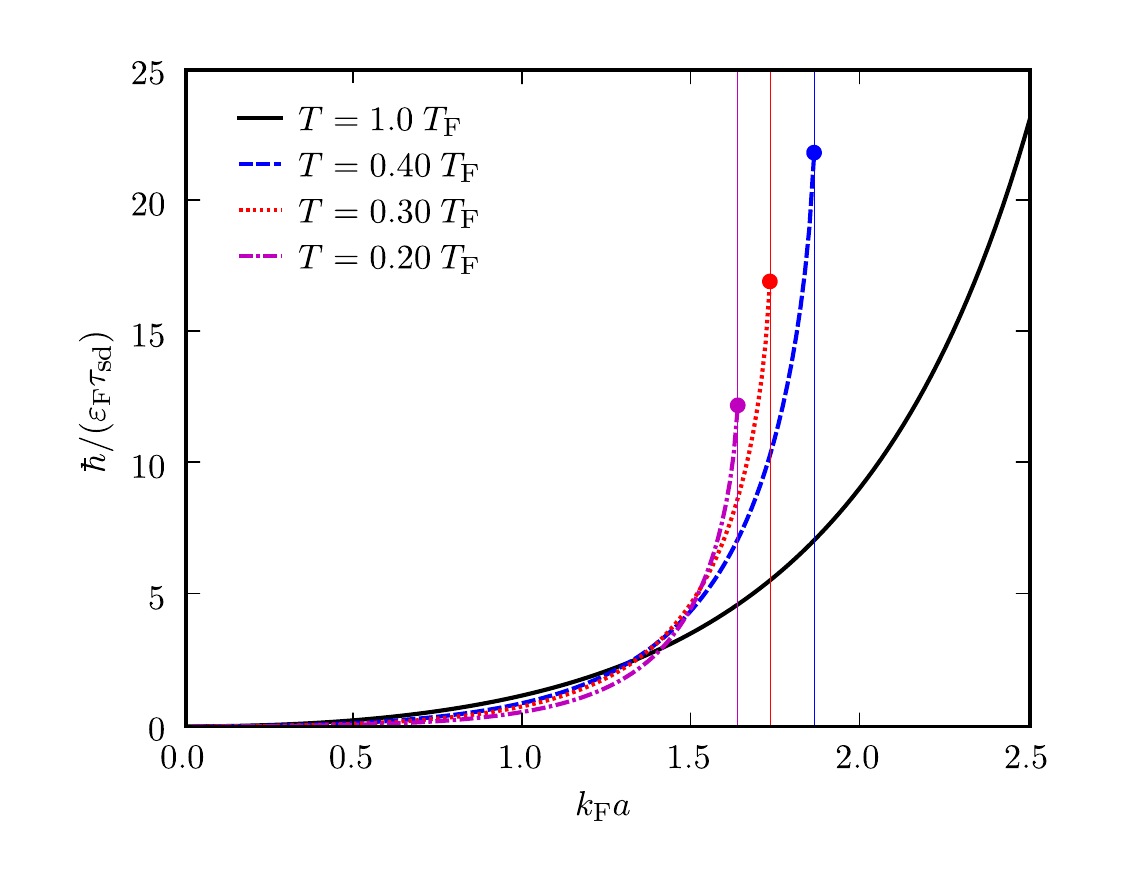}
\caption{(Color online) Spin drag relaxation rate $1/\tau_{\rm
sd}$ as a function of $k_{\rm F} a$ for various temperatures. The
thin vertical lines indicate the critical values of the
interaction parameter $k_{\rm F} a$ at which ferromagnetism
occurs.\label{fig:tausdvskfa}}
\end{center}
\end{figure}

\begin{figure}
\begin{center}
\includegraphics[width=1.00\linewidth]{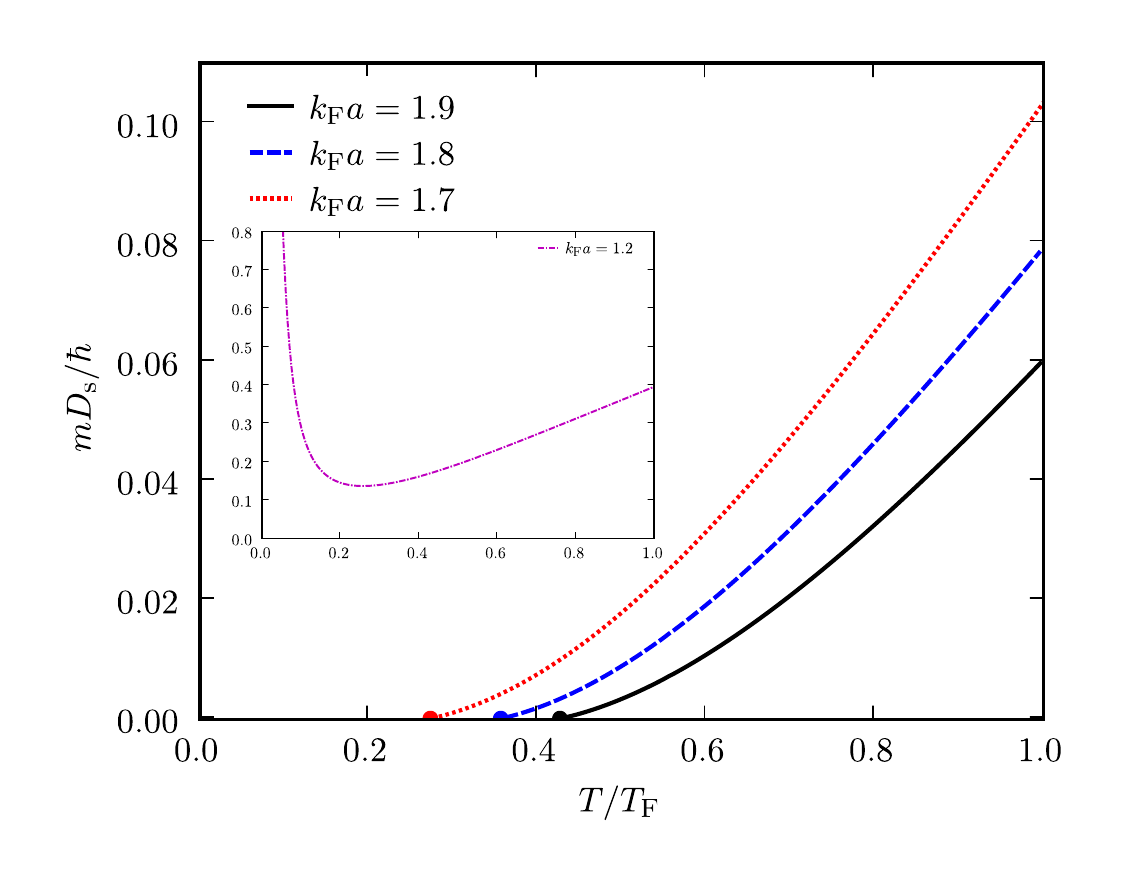}
\caption{(Color online) Panel a): spin diffusion constant $D_{\rm
s}$ as a function of temperature $T$ for various interaction
strengths $k_{\rm F} a > \pi/2$. Panel b): spin diffusion
constant for $k_{\rm F}a =1.2 < \pi/2$.\label{fig:DvsT}}
\end{center}
\end{figure}

{\it Results.} --- In
Figs.~\ref{fig:tausdvsT}~and~\ref{fig:tausdvskfa} we present the
results of a numerical evaluation of Eq.~(\ref{eq:resulttausd}),
both as a function of temperature for various interaction
strengths $k_{\rm F}a$, and as a function of interaction strength
for various temperatures. In experiments both dependencies can be
explored using a Feshbach resonance to tune the interaction
strength \cite{feshbach}. Both figures clearly show the strong
enhancement of the spin drag relaxation rate as the ferromagnetic
state is approached. The precise form of the enhancement is
understood by keeping only the most divergent term in the
scattering amplitude in
Eq.~(\ref{eq:scattering_amplitude_transparent}) so that
\begin{equation} \label{eq:Aapprox}
 A_{\uparrow\downarrow} (q,\omega)
 \simeq - \frac{3 U^2 \nu (\varepsilon_{\rm F})}{\displaystyle \frac{U \nu(\varepsilon_{\rm F})}{3} \left(
 \frac{q^2}{k_{\rm F}^2}- i
\frac{6\pi m \omega}{\hbar k_{\rm F} q} \right)+4 \alpha (T)}~,
\end{equation}
with $\alpha (T) = 1 + U \chi_0(0,0)/2 \simeq 1 - U\nu
(\varepsilon_{\rm F}) + \pi (T/T_{\rm F})^2/12 + \dots$ equal to zero at
the phase transition and positive for $T$ larger than the critical
temperature. In obtaining this approximation we have expanded
around the ferromagnetic singularity and used that $ \Im
m~\chi_0(q,\omega) = - \pi \nu (\varepsilon_{\rm F}) m \omega/2\hbar
k_{\rm F} q$. Using these results in Eq.~(\ref{eq:resulttausd}),
and expanding $1/\sinh^2(x) \simeq 1/x^2$, both the frequency
integral and the momentum integral can be performed analytically
if we use a cut-off of $2k_{\rm F}$ on the momentum integration
that diverges because of the expansion of $1/\sinh^2(x)$.
Ultimately we find in this manner that $1/\tau_{\rm sd} (T) -
1/\tau_{\rm sd} (T_{\rm c}) \propto (T-T_{\rm c})\ln(T-T_{\rm c})$ for
$T \downarrow T_{\rm c} \ll T_{\rm F}$, which indeed accurately
describes our numerical results near the critical temperature.

We also consider the spin diffusion constant, which from the
Einstein relation is given by $D_{\rm s}(T) = \sigma_{\rm s}(T)/
\chi_{S_zS_z}(0,0)$ and the ``spin conductivity" $\sigma_{\rm s}(T) = n
\tau_{\rm sd}(T)/ m$~\cite{giovanni_cluj_2007}. In
Fig.~\ref{fig:DvsT} we show this constant as a function of
temperature for various values of the interaction strength. Near
the critical temperature the spin diffusion constant vanishes as
$D_{\rm s}(T) \propto (T-T_{\rm c})^\kappa$ with an exponent
$\kappa =1$, because $\tau_{\rm sd} (T_{\rm c})$ remains finite
and $\chi_{S_zS_z}(0,0)$ diverges as $1/(T-T_{\rm c})$ within our
random-phase approximation.

At this point it is important to
realize that from the point of view of critical dynamics our
findings are mean-field like. If the spin dynamics can
be effectively described by an isotropic Heisenberg ferromagnet
(model J \cite{hh_review}), the spin conductivity $\sigma_{\rm s}(T)$ is expected to
behave as $\xi^{(3-\eta)/2}$ very close to the transition, where
the correlation length $\xi(T)$ diverges as $1/(T-T_{\rm c})^\nu$ and
$\eta$ and $\nu$ are the usual static critical exponents of the
ferromagnetic transition \cite{hh_review}. Since
$\chi_{S_zS_z}(0,0)$ diverges as $\xi^{2-\eta}$, we find that the
spin diffusion constant goes to zero with an exponent $\kappa =
(1-\eta)\nu/2$. In view of this possibility, we have therefore
made sure, using the Ginzburg criterion~\cite{ginzburg}, that the
upturn of the spin relaxation rate takes already place well
outside the critical region where critical fluctuations can be
neglected and our random-phase approximation is appropriate.

{\it Discussion and conclusions} --- As we have mentioned in the
introduction, the spin drag relaxation rate can be determined from
the damping of the spin dipole
mode~\cite{bruun_prl_2008,damico_prb_2006} in a trapped gas. Since
the Fermi energy is usually much larger than the level splitting
in the trap, our results show that the spin dipole mode is
typically strongly overdamped, which makes the experiment more
challenging. Nevertheless, such a measurement, as well as
measurements of the spin diffusion constant as a function of
temperature, gives information on the proximity of the
ferromagnetic phase transition. Although we have considered a
homogeneous system, local-density approximations are generally
valid for trapped Fermi systems, and in determining the damping of
the spin dipole mode of a trapped two-component gas the
homogeneous density should in first approximation be taken as the
central density of the atomic cloud.

Within our present approach, we consider the transition to
ferromagnetism within MFT, which predicts it to be continuous.
One interesting aspect is that taking into account correlation effects beyond MFT~\cite{duine2005} results in: (i) an increase in the critical temperature for a given value of $k_{\rm F}a$ and (ii) a change in the character of the transition from second to first order at very low temperatures. We expect that these modifications will not qualitatively affect the upturn of $1/\tau_{\rm sd}$, as long as one remains outside the critical-fluctuation region.

In future work we intend to explore the effects of lower
dimensionality, and the implications of critical and quantum
critical fluctuations on the exponent $\kappa$ that determines the
behavior of the spin diffusion constant. Furthermore, as our
approach is distinct from the work of Hu~\cite{hu2000}, who
considered the upturn of the Coulomb drag resistivity as one
approaches the exciton-condensed state in bilayers, we intend to
study this system as well with our present approach.

\acknowledgements This work was supported by the Stichting voor
Fundamenteel Onderzoek der Materie (FOM), the Netherlands
Organization for Scientific Research (NWO), and by the European
Research Council (ERC). G.V. acknowledges support from NSF Grant
No. 0705460. M.P. acknowledges very useful conversations with
Rosario Fazio and Andrea Tomadin.

\end{document}